# A scale-free method for testing the proportionality of branch lengths between two phylogenetic trees

Yichen Zheng[1], William Ott[2], Chinmaya Gupta[2] and Dan Graur[1]

1 Department of Biology and Biochemistry, University of Houston, Texas 77204-5001, USA

2 Department of Mathematics, University of Houston, Texas 77204-3008, USA

Email addresses: Yichen Zheng: yzheng7@uh.edu; William Ott: ott@math.uh.edu; Chinmaya Gupta: chinmaya@math.uh.edu; Dan Graur: dgraur@uh.edu

Corresponding author: Yichen Zheng

*Abstract*

We introduce a scale-free method for testing the proportionality of branch lengths between two phylogenetic trees that have the same topology and contain the same set of taxa. This method scales both trees to a total length of 1 and sums up the differences for each branch. Compared to previous methods, ours yields a fully symmetrical score that measures proportionality without being affected by scale. We call this score the normalized tree distance (NTD). Based on real data, we demonstrate that NTD scores are distributed unimodally, in a manner similar to a log-normal distribution. The NTD score can be used to, for example, detect co-evolutionary processes and measure the accuracy of branch length estimates.

*Introduction*

In species groups with established phylogenetic relationships, such as the great apes, branch length is used to compare the rates and patterns of evolution among different lineages and among different genes. Let us consider a case of two orthologous groups of genes from the same taxa. On each branch, gene *a* will evolve at a rate determined by the rate of mutation on that branch and the selective constraints that are dictated by its function. The same applies to gene *b* that performs a different function. If the lineages under study experience mutation rates that do not change with time (but vary between genes) and if the two genes maintain their respective functions in the lineages under study, then the branch lengths of the phylogenetic tree for gene *a* (tree ***A***) will most probably be different from the branch lengths of the phylogenetic tree for gene *b* (tree ***B***); however, the corresponding branches on the two different trees will be proportional to each other. That is, dividing the length of a branch in tree ***A*** by the length of the corresponding branch in tree ***B*** will yield the same result regardless of which branch pair is chosen (Figure 1a, b). If, on the other hand, the selective constraints or the mutation patterns change in one or more branches, proportionality will be violated (Figure 1a, c).

Methods for comparing phylogenetic trees, especially branch lengths, can be used in studying the patterns of molecular evolution. For example, Pazos et al. (2008) compared phylogenetic trees of bacterial proteins and found that tree similarity can be predictive for protein interaction. Lovell and Robertson (2010) also suggested that the similarity of branch length ratios, or "evolution rate correlation," is an indicator of protein-protein interactions. Rosa et al. (2013) used branch length comparison as one of the methods to characterize the evolution of "barcode sequences" (stretches of mitochondrial DNA used to identify species.) When determining the accuracy of phylogenetic tree reconstruction, the accuracy of branch lengths is an important aspect to consider (e.g., García-Pereira et al. 2011, Knowles et al. 2012).

Measures of proportionality between two trees should be free of bias caused by the scale of the two trees. Mathematically, there are two issues that must be dealt with. First, the distance between two trees should be independent of scale: resizing one (or both) of the trees by multiplying all branch lengths by a fixed number should leave the distance between the trees unchanged. For example, since trees ***A*** and ***B*** in Figure 1a and 1b are different in scale but perfectly proportional to each other, the distance between them should be zero (or their "similarity" should be 100%). Second, the distance function should be a metric in the mathematical sense, meaning that it should be symmetric and satisfy the triangle inequality. The triangle inequality implies that the distance between trees ***A*** and ***C*** in Figure 1 should be equal to that between trees ***B*** and ***C***. Such a scale-independent, mathematically rigorous notion of distance would be useful in a variety of contexts. In particular, scale independence prevents bias due to longer trees appearing to have larger distances from one another.

There are a few methods in the literature that compare phylogenetic trees; however, most of them only take differences in topology into account (e.g., Robinson and Foulds 1981; Nye et al. 2006). One of the very few distance measures that take both topology and branch length into consideration is the Branch Length Score (BLS) by Kuhner and Felsenstein (1994):

$$BLS = \sqrt{\sum (a_i - b_i)^2} \qquad (1)$$

Here $a_i$ and $b_i$ are the branch lengths corresponding to the *i*-th possible bipartition of all the taxa in trees **A** and **B**, respectively. This measure is implemented in the popular phylogenetic package PHYLIP (Felsenstein 2005). The BLS measure, however, depends on scale: trees with longer branch lengths will produce larger BLS values. In addition, a large BLS value will be produced if the trees are proportional to each other but the rates of evolution are different. For example, the BLS between trees **A** and **B** in Figure 1 is 0.0255, while the BLS between trees **B** and **C** is 0.0197. Therefore, BLS can be affected by non-lineage-specific variation of evolution rate, i.e., tree scale, which makes it an inappropriate measurement of proportionality.

To counter this problem, Soria-Carrasco et al. (2007) made an ingenious modification to BLS by scaling one of the trees with a parameter K that minimizes BLS. This modified distance measure is called the "K tree Score" (KTS).

$$KTS = \sqrt{\sum_{i=1}^{N}(a_i - Kb_i)^2}, where\ K = \frac{\sum_{i=1}^{N}(a_i b_i)}{\sum_{i=1}^{N} b_i^2} \tag{2}$$

This value of K is chosen to minimize the score. Because only one of the trees is scaled, this measure is not symmetrical. Here the capital "*N*" is used to denote the total number of branches, as opposed to the lower-case "*n*" which will be used to represent number of species.

Another tree-comparing algorithm that uses branch length data is Hall's CompareTrees program (as used in Hall 2005). Unlike BLS, one of the two trees has to be designated the "true tree." This branch length score (CompareTrees Score, CTS) is calculated by averaging the relative differences between the lengths of the same branches in the two trees:

$$CTS = (\sum_{i=1}^{N} 1 - \frac{|a_i - b_i|}{a_i})/N \tag{3}$$

Here $N$ is the number of branches shared by the two trees and $a_i$ and $b_i$ are the lengths of the shared branches. In this method, *A* is designated the "true tree" or reference tree, and *B* is the tree compared to it. This method is intended to be used when one of the trees is known to be true. Similar to KTS, this method is asymmetrical. Furthermore, if there is a very short branch in *A*, it may produce an extremely large value due to being a denominator; this may obscure the comparisons of other branches.

*The NTD method*

Here, we propose a method for comparing phylogenetic trees that solves the mathematical challenges outlined above. The method uses what we call the normalized tree distance (NTD) and is suitable for comparing trees with the same topology and set of taxa.

Imagine two unrooted phylogenetic trees, *A* and *B*, with the same topology and the same set of $n$ taxa. Since the topology is identical, each tree can be described by $N = 2n-3$ branch lengths. They are denoted by $a_1, a_2, a_3,\ldots, a_N$ and $b_1, b_2, b_3,\ldots, b_N$. As a consequence, each phylogenetic tree is represented by vector in an $N$-dimensional space: $A = (a_1, a_2, \ldots, a_N)$, and $B = (b_1, b_2, \ldots, b_N)$. Comparing the trees can be done by comparing the two vectors. The measure we choose, the NTD, is derived by adding up numerical differences between each pair of branches after both trees are scaled to a total branch length of 1, then dividing the sum by 2:

$$NTD = \left(\sum\left(\left|\frac{a_i}{\sum_{j=1}^{N} a_j} - \frac{b_i}{\sum_{j=1}^{N} b_j}\right|\right)\right)/2 \tag{4}$$

As the added differences are all absolute values, the NTD will always be greater than or equal to 0. At the same time, the theoretical maximal value is 1; this happens when all branches with non-zero length in tree *A* have zero length in tree *B* and vice versa. In this situation, the differences will add to 2; after dividing by 2, the NTD will be 1. The range [0,1] of the NTD does not change with either number of taxa or the total length of the trees; therefore, this measure is fully normalized.

In mathematical terms, the calculation of NTD after scaling is the L1 metric on the set of N-vectors whose nonnegative entries sum to 1. Like all mathematical metrics, NTD is therefore symmetric and satisfies the triangle inequality.

*Simulated Example*

Let us first compare our NTD with scores obtained by the three other methods: BLS, KTS and CTS. Two eight-taxon phylogenetic trees (Figure 2) with the same topology but different branch lengths were randomly generated and named *X1* and *Y1*. All branch lengths in tree *Y1* were doubled to produce tree *Y2*, and tripled to produce tree *Y3*. Here we will examine the properties of different tree-comparing methods using these simulated phylogenetic trees.

Table 1 shows the scores given by all four methods. From the comparison between *X1* and *Y1* and between *Y1* and *X1*, it is clear that both KTS and CTS produce asymmetrical results. From the comparison between *Y1* and *Y2*, we see that only NTD and KTS recognize perfectly proportional trees. Also, when one of the compared trees has very long branches (e.g., when *X1* and *Y3* are compared), the BLS and the CTS will have large absolute values, while NTD is always between 0 and 1. No matter in which order are they compared or if the branches are proportionally changed, as long as the comparison is done between an "*X*" and a "*Y*" tree, NTD will remain the same.

*Deriving distribution of NTD*

Here we will provide a profile of how the distribution of NTD looks from real DNA and protein sequence data.

We downloaded CDS and corresponding protein alignments that contain 12 well-sequenced mammal species from the online database ORTHOMAM (Douzery et al. 2014). These species are: human (*Homo*), chimpanzee (*Pan*), macaque (*Macaca*), marmoset (*Callithrix*), rat (*Rattus*), mouse (*Mus*), guinea pig (*Cavia*), dog (*Canis*), horse (*Equus*), cow (*Bos*), pig (*Sus*) and elephant (*Loxodonta*). The website also provides the topological phylogenetic relationship among these species (Figure 3). Although there is controversy on the placement of horse (e.g., Zhou et al. 2012), we decided to use this external tree as the user tree for simplicity. All alignments containing unknown nucleotides or amino acids were removed. 5,140 pairs of DNA/protein alignments remained in the dataset (see Supplementary Material for a list of the genes).

All CDS and protein alignments were used to produce maximum likelihood trees with RAxML (Stamatakis 2006), using the GTRGAMMA model for DNA sequences and PROTGAMMAJTT model for protein sequences. A user tree (Figure 3) was used to guide the tree topology. Branch lengths were collected from the result for calculating NTD. Three empirical distributions were derived: among all DNA trees, among all protein trees, and between DNA and protein trees for each gene. The distributions were fit to beta, gamma and lognormal distributions, using log likelihood as a measure of goodness-of-fit.

Distributions were fit to the NTD data by using a maximum likelihood estimation of the distributions parameters, together with resampling to ensure that the fitted distributions are robust to outliers in the given data.

For a given probability distribution with probability density function $f$, the log likelihood (LL) of data $X_1, \ldots, X_M$, given a parameter $\theta$, is computed as $\text{LL}(\theta) = \sum_{i=1}^{M} \log f(X_i|\theta)$. We normalized this log likelihood by the sample size to better compare the LL for data sets of different sizes. Therefore, instead of using LL($\theta$), we used L($\theta$) := LL($\theta$)/$M$.

To fit distributions to the NTD data, we used maximum likelihood estimation, together with a resampling technique to ensure that the fitted distributions are robust against outliers in the datasets. Given the original dataset $X_1, \ldots, X_P$, a family of subsamples $X_1^j, \ldots, X_{P_j}^j$ ($j = 1,\ldots,$ 1000, $P_j$ = 1000 for all $j$) was taken. For each subsample, the maximum likelihood estimate for the unknown parameter $\theta$ was computed. We call this estimate $\theta_j$. We then used a resampling technique to compute a log likelihood value for each $\theta_j$. Specifically, from the original data we then sampled $k$ = 1000 subsamples, each of size K = 1000, and used these subsamples to obtain repeated estimates $\{L(\theta_j)_i\}_{i=1}^{k}$. The log likelihood for the parameter $\theta_j$ was obtained by taking the mean $\hat{L}(\theta_j) = \frac{1}{k}\sum_{i=1}^{k} L(\theta_j)_i$. Finally, we chose the best fit estimator $\hat{\theta}$ as

$$\hat{\theta} = \arg\max_{\theta_j} \hat{L}(\theta_j).$$

The error on the estimate $\hat{L}(\hat{\theta})$ was obtained by computing $\sigma/\sqrt{k}$ where $\sigma$ is the sample standard deviation of $\{L(\hat{\theta})_i\}_{i=1}^{k}$.

*Results*

Figure 4 shows empirical distributions of NTD score in three different datasets and approximation of them to established distribution families. Although they are all unimodally distributed, we see that the mean score for comparisons between corresponding DNA and protein trees (Figure 4c) is smaller than the mean score for comparisons among DNA trees (Figure 4a), which is in turn smaller than the mean score for comparisons among protein trees (Figure 4b). This can be explained biologically, as DNA sequences have more neutral-evolving characters than proteins and the lineage-specific selection effects are weaker. Trees produced from the DNA and protein alignments of the same gene are more similar because these two sources are dependent on each other.

Table 2 shows the log likelihood scores for distribution fitting. For all three datasets, lognormal distribution fits better than both beta and gamma distributions, though the scores are not highly different. Similarly, in Figure 4, one can see that the green curves (lognormal) fit the histograms better than the black (beta) and red (gamma) curves. We suggest that the NTD scores of a sample of phylogenetic trees produced from real sequences are distributed most closely to lognormal.

*Discussion*

Differences between phylogenetic trees can be classified into three categories: differences in topology, scale, or proportionality. While most comparison methods (Robinson and Foulds 1981; Nye et al. 2006) focus on topology, the ones that compare branch lengths do not distinguish between scale (Kuhner and Felsenstein 1994) and proportionality, or are not symmetrical (Soria-Carrasco et al. 2007, Hall 2005). Our NTD measure is useful when only proportionality information (but not scale) is needed.

For the NTD method, we chose to use the total tree length to scale each tree to total length one before comparison. If the tree is considered a vector, the scaling factor is known as the L1-norm ($a'_i = a_i / \sum_{j=1}^{N} a_j$). In mathematical algorithms that deal with vectors, however, a popular scaling alternative is the L2-norm ($a'_i = a_i / \sqrt{\sum_{j=1}^{N} a^2}$), which is the square root of the sum of the squared values. We compared the L1-scaled trees branch by branch and used the sum of the differences, known as the L1-distance ($\sum_{i=1}^{N}(|a'_i - b'_i|)$). We certainly could have used the L2-distance (Euclidean distance, $\sqrt{\sum_{i=1}^{N}(a'_i - b'_i)^2}$), since Euclidean distance has a natural geometric meaning. The main reason we chose L1 measures over L2 is the consideration of the biological meaning of branch lengths. Since branch length signifies the amount of evolutionary change from one point in the tree to another, the total length (L1-norm) is the total amount of evolutionary change that occurred along the entire tree, while the L2-norm does not have a biological meaning.

Unlike previous methods, the NTD is itself normalized, because both trees are scaled before the comparison. All possible NTD scores are between 0 and 1. This enables the NTD to become a standardized measure of tree proportionality, which can be compared across different taxa. Another advantage of this method is that it does not need too much computation time. To calculate the NTD between two *n*-taxa trees, only 8*n* – 15 additions/deductions and 4*n* – 5 divisions are needed. No iterations are required, and the computation time increases linearly with the number of taxa.

The potential applications for such a measure include comparing the evolutionary histories of two proteins where the phylogeny among the species studied is uncontroversial or known (e.g., experimentally evolved organisms). For example, if the NTD between gene *a* and gene *b* is much lower than those between gene *a* and other genes, it is possible that genes *a* and *b* coevolve. In addition, because the absolute values of branch lengths are not important, NTD can be used to compare trees computed from different kinds of data, e.g., from DNA sequences and protein sequences. Even if the absolute branch-length values in a DNA tree and a protein tree are not comparable directly, the trees can be compared using NTD. If comparing phylogenetic trees produced by a coding gene and its protein product gives a high NTD score, it is likely that the selection pressure on this gene differs among lineages.

As the NTD is a global measure that compares entire trees, it can only be used to compare the branch length patterns over the entire trees. To pinpoint in which lineages the selective changes occur, lineage-specific measures such as dN/dS analysis are required. However, since NTD is easy and fast to compute, it is computationally efficient to first identify "interesting" tree pairs with NTD before analyzing them in depth with more sophisticated methods like maximum likelihood or Bayesian analysis. In the future, a statistical test can be devised that uses NTD scores to identify genes that evolve under different situations in a single set of species. For example, researchers can establish a lognormal distribution for one-to-one comparisons in a collection of gene trees, and look for trees that produce significantly more scores in the tail of the distribution compared to the mean.

Finally, we want to mention the possibility of using NTD for trees that are not topologically identical. In the BLS method (Kuhner and Felsenstein 1994), branches that are present in one tree but not in the other are treated as zero length in the latter tree. This can also work in our NTD measure; however, we decided not to include this aspect here, since we are skeptical as to how well a zero-length branch can represent a non-existent one in a phylogenetic tree.

Figure 1 Phylogenetic trees for three hypothetical proteins, *a*, *b,* and *c*. Between *a* and *b*, there are no lineage-specific changes in the selection scheme, so the branch lengths are completely proportional; however in protein *c*, the selection has strengthened or relaxed in some branches, causing a deviation from proportionality.

Figure 2 The simulated 8-taxa trees used to compare the four measures. ***X1*** and ***Y1*** are produced independently, while ***Y2*** and ***Y3*** are produced respectively by doubling and tripling each branch length of ***Y1***.

Figure 3 The phylogenetic tree topology of 12 mammalian species. This is a commonly accepted topology of their phylogenetic relationship, used as a guide tree for the maximum likelihood tree reconstruction.

Figure 4 The distribution of NTD scores from a sample of mammalian gene trees. (a) Trees produced from DNA sequences are compared with one another; (b) trees produced from protein sequences are compared with one another; (c) trees produced from corresponding DNA and protein sequences are compared with each other for each gene. The curves are theoretical distributions fit to the data with a maximum likelihood method. The black curves represent the best-fit beta distribution, green curves the best-fit lognormal distribution, and red curves the best-fit gamma distribution. In all three cases, the lognormal distribution fits most appropriately.

Table 1. Comparison of four tree-comparing measures used simulated eight-taxon trees (Figure 2). These scores are, the Normalized Tree Distance (NTD), the branch length score (BLS), the K tree score (KTS), and the CompareTrees score (CTS).

| Tree Pairs | NTD | BLS | KTS | CTS |
| --- | --- | --- | --- | --- |
| *X1 / X1* | 0 | 0 | 0 | 1 |
| *Y1 / X1* | 0.08144 | 2.3054 | 1.71401 | -1.09598 |
| *X1 / Y1* | 0.08144 | 2.3054 | 2.18746 | -0.34057 |
| *X1 / Y2* | 0.08144 | 3.47463 | 2.18746 | -1.63611 |
| *X1 / Y3* | 0.08144 | 5.15815 | 2.18746 | -3.00397 |
| *Y3 / X1* | 0.08144 | 5.15815 | 5.14204 | 0.27316 |
| *Y1 / Y2* | 0 | 1.97171 | 0 | 0 |
| *Y2 / Y1* | 0 | 1.97171 | 0 | 0.5 |

Table 2. The log-likelihood of three distribution families fit to the NTD data from protein-protein, DNA-DNA and protein-DNA tree comparisons. In all cases, lognormal distribution appears to have the highest log-likelihood values, indicating a good fit. Error estimates are in parentheses.

| Dataset | Beta distribution | Lognormal distribution | Gamma distribution |
| --- | --- | --- | --- |
| Protein-protein | 0.3612 (<0.0001) | 0.4482 (0.0004) | 0.4372 (0.0004) |
| DNA-DNA | 1.0107 (0.0011) | 1.0644 (0.0009) | 1.0447 (0.0008) |
| Protein-DNA pairs | 0.9369 (0.0010) | 1.0515 (0.0009) | 0.9918 (0.0009) |

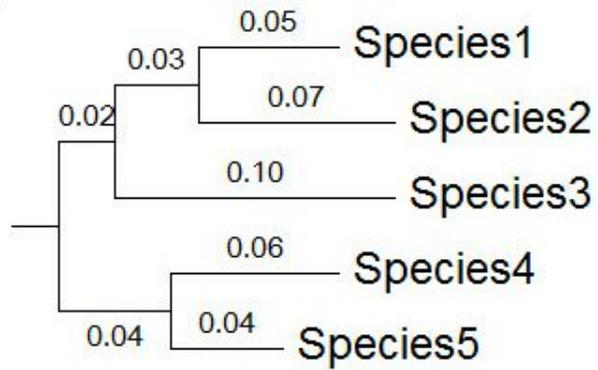

Tree *A*

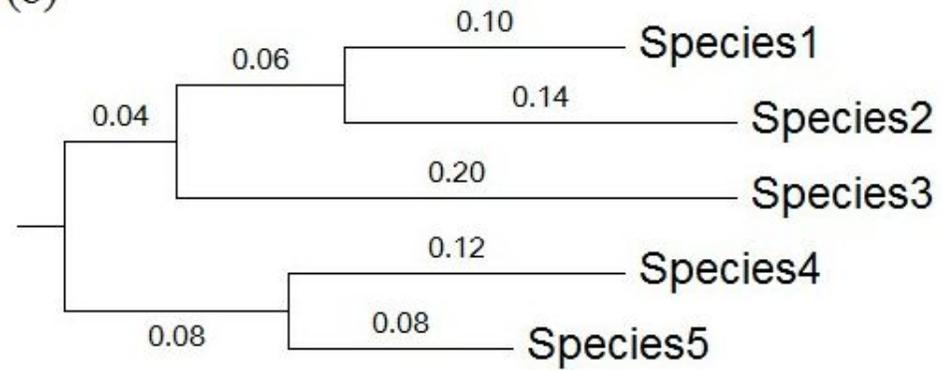

Tree *B*

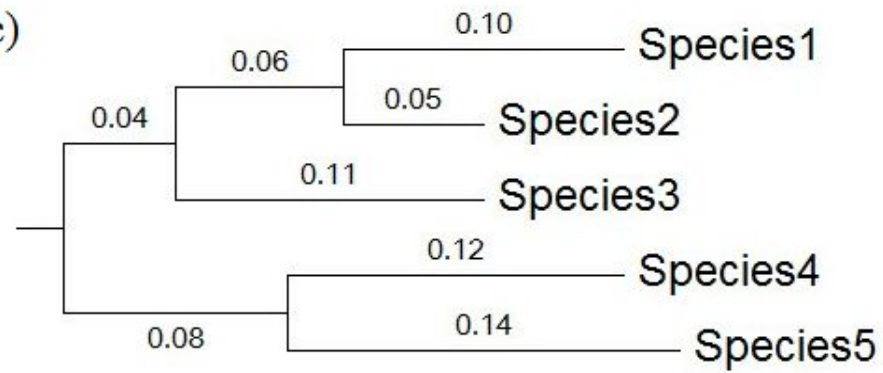

Tree *C*

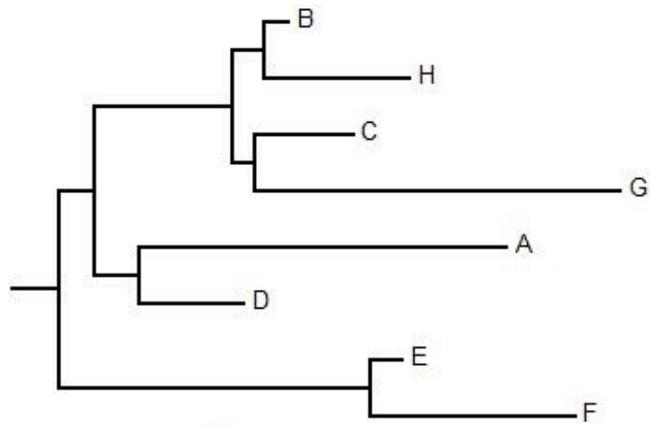
X1

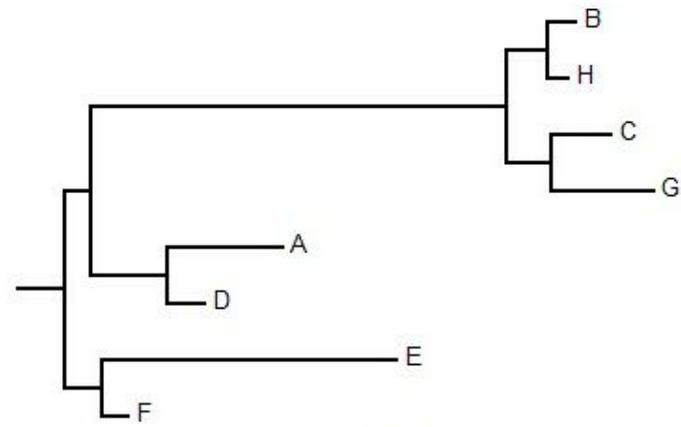
Y1

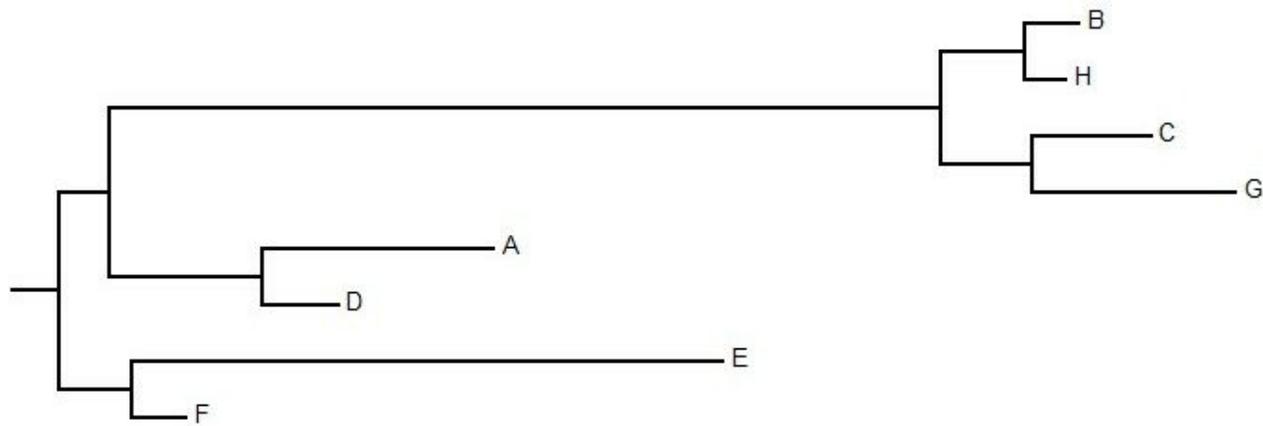
Y2

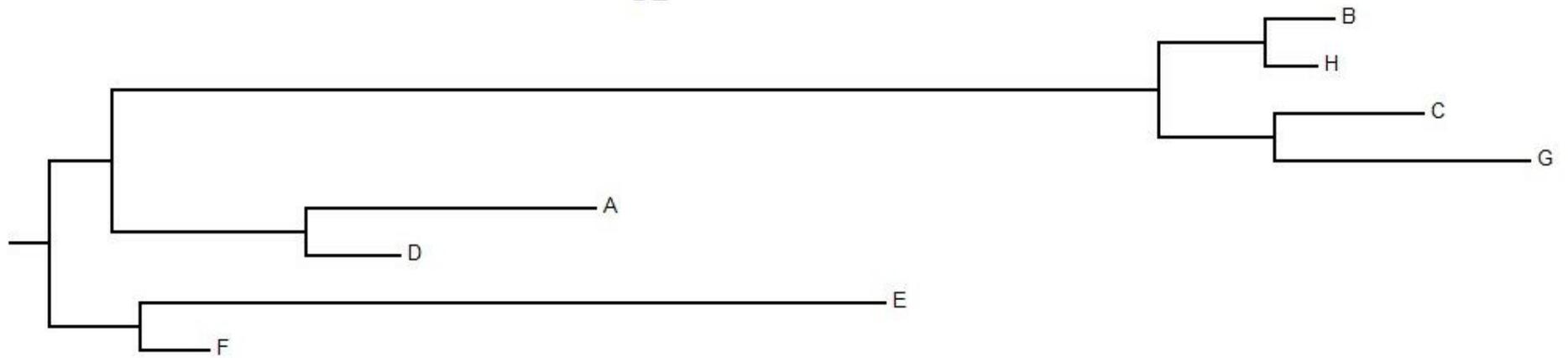
Y3

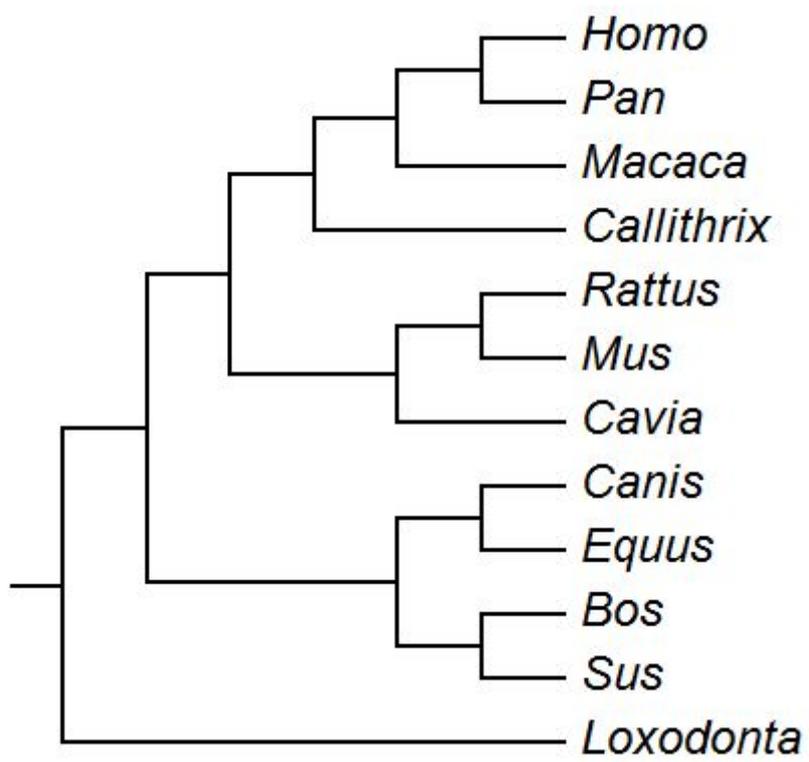

(a) 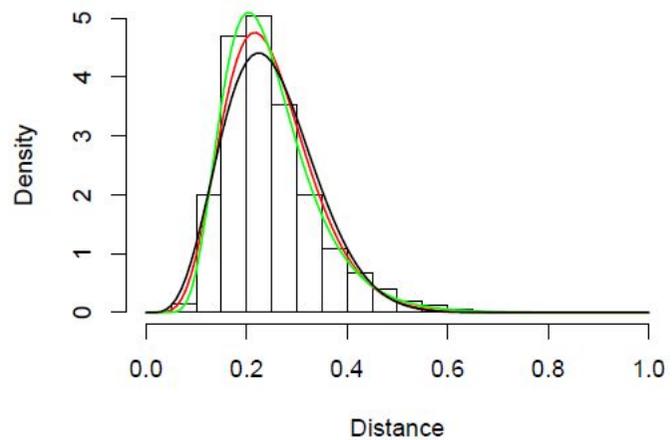
(b) 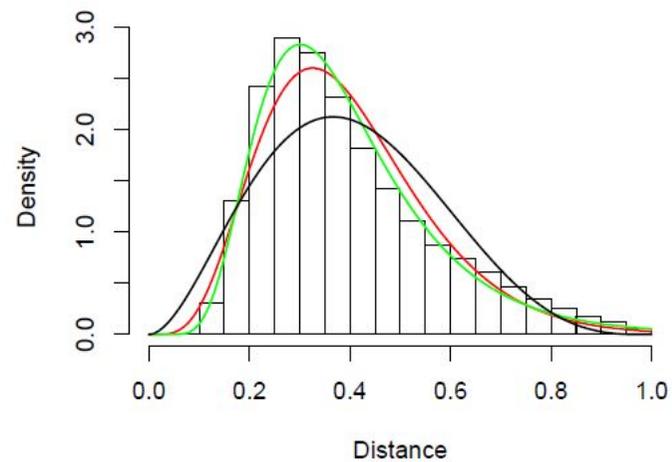
(c) 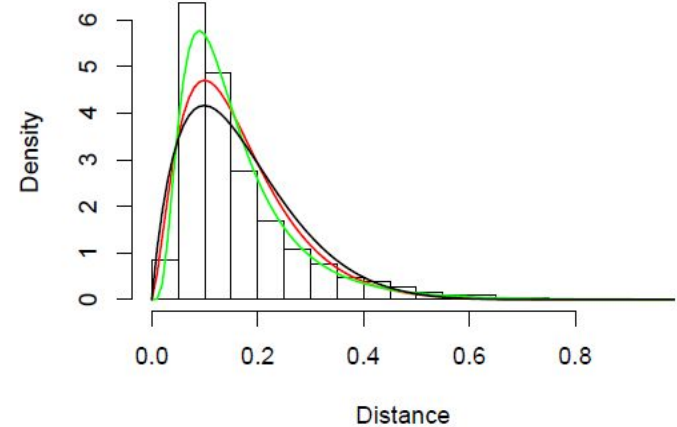